\documentclass[12pt,preprint,a4paper]{aastex}
\usepackage{times}
\usepackage{graphics,epsf}
\usepackage{amsmath}                % American Mathematical Society package
\usepackage{amsfonts}               % American Mathematical Society fonts
\usepackage{amssymb}                % American Mathematical Society symbol
\usepackage{epsfig}                 % EPS figures
\usepackage{graphicx}
\usepackage{url}
\usepackage{tabularx}
\usepackage{booktabs}
\usepackage{hyperref}
\usepackage[dvipsnames]{xcolor}
\def \s{~\rm{s}}

\def \km{~\rm{km}}

\def \AU{~\rm{AU}}

\begin{document}

\title{A summary of two meetings on the role of jets in interacting stellar binaries}
\author{Noam Soker}
\affil{Department of Physics, Technion -- Israel Institute of
Technology, Haifa 32000, Israel;
soker@physics.technion.ac.il}

\begin{abstract}
In relation to two meetings on stellar binary systems in July 2017, I summarize my view that in the majority of strongly interacting stellar binary systems jets play decisive roles. In the meeting \emph{{The Physics of Evolved Stars II: The role of Binarity}} (Nice, July 10-13, 2017) many talks and posters were concentrated on the shaping of circumstellar matter, with some significant new results, like the observations of wide jets launched from binary systems. In the meeting \emph{The Impact of Binaries on Stellar Evolution} (ESO Garching, July 3-7, 2017), both in the opening and the summary talks the speakers mentioned that ``Textbooks need most likely to be rewritten''. I made my point that some new chapters were already written, although researchers do not read them but rather keep referring to the `old textbooks'. For example, too many speakers and authors keep referring only to the single degenerate and double degenerate scenarios for the progenitors of SNe Ia. Researchers must refer also to the `new textbook chapter' on the core degenerate (CD) scenario. Other examples involve the role of jets. Jets lead to a new evolutionary phase, that of the grazing envelope evolution (GEE), that should be considered alongside the older ones, e.g., the common envelope evolution, Roche lobe overflow (RLOF), and wind mass loss and accretion. I discuss how the GEE can solve some puzzling binary systems. We should also consider the role of jets as an extra energy source to remove the common envelope and in powering intermediate luminosity optical transients (ILOTs). One strong advantage of jets is that in many cases they operate through a negative feedback mechanism, hence preventing the need for a fine tuning. The operation of the jet feedback mechanism (JFM) connects also binary evolution to explosion of massive stars as core collapse supernovae.
\end{abstract}

% ==========================================================
\section{INTRODUCTION}
\label{sec:introduction}
% ==========================================================
% =======================
\subsection{Preface}
\label{subsec:preface}
% =======================
I summarize the talks and posters on the subject of interacting binary systems that were presented during the meetings \emph{The Impact of Binaries on Stellar Evolution} (ESO, Garching, July 3-7, 2017\footnote{The link to the meeting:
{http://www.eso.org/sci/meetings/2017/Imbase2017.html}}) and \emph{{The Physics of Evolved Stars II: The role of Binarity}} (Nice, July 10-13, 2017\footnote{The link to the meeting:
{https://poe2017.sciencesconf.org/}}). This is not a review article on the subject, hence I do not list many relevant references, and I limit myself to the talks and posters that were presented at the meetings, and only to those in relation to strongly interacting binary systems.
As well, this is not an official summary of the meetings. The official summary of the ESO meeting was given by Ed van den Heuvel and is on the site of the meeting (as are the other talks; in  `Program'). Nathan Smith gave the official summary of the meeting in Nice.

In his final talk of the ESO meeting, Ed van den Heuvel quoted a line he took from the first talk of the ESO meeting, given by Henri Boffin: ``textbooks need most likely to be rewritten!''
As I will discuss throughout the manuscript, some chapters of the `new textbook' have been already written, but most speakers overlooked them.

In the meeting in Nice we had plenty of discussion time. In the meeting in ESO, on the other hand, there was not even one discussion period. Already in the first talk by Henri Boffin only two people had the opportunity to comment; many others raised their hands, but there was no time for more questions and comments (in some other cases there was more time, and good comments were made in the question periods, e.g., by Zhanwen Han in a number of cases). As such, I presented a live-poster, where twice a day I updated it with a summary and personal comments to the relevant talks and posters of that day. Many parts of this manuscript are based on my live poster.
To view my live poster go to the date of July 2017 in the link:
https://phsites.technion.ac.il/soker/category/news/

% =======================
\subsection{Strongly interacting binary systems}
\label{subsec:strongly}
% =======================
By strongly interacting binary systems I refer to evolutionary routes where at least one of the properties of at least one star in the binary system is determined by the other star.
I list several examples for such properties.

\emph{Outflow kinetic energy.} Evolved stars have massive winds. The energy per unit mass of the wind of an asymptotic giant branch (AGB) star is $\approx (10 \km \s^{-1})^2$. Consider a main sequence companion outside the envelope of the giant star that accretes mass through an accretion disk, and launches jets that carry a fraction of $\approx 0.1$ of the accreted mass. The jets' velocity is about the escape velocity, with an energy per unit mass of $\approx (400 \km \s^{-1})^2$. For these parameters it is sufficient that the main sequence companion accretes $\approx 1 \%$ of the AGB wind to overpower the AGB wind.
Even if the energy in the jets is only, say, $\approx 10\%$ of the AGB wind, I still consider the system as strongly interacting because along the polar directions the jets overpower the AGB wind.
This argument goes to other types of binary systems that experience a similar wind-jets interaction.

\emph{Radiated energy.} The radiation that is emitted by the accreted gas onto the compact companion, and/or the radiation that is emitted as a result of the interaction of the jets with the wind, can overpower the combined luminosity of the two stars that results from nuclear burning. Such binary systems are strongly interacting.

\emph{Envelope energy.} A compact companion that enters the envelope of a larger star deposits orbital energy to the envelope. The companion can also launch jets that energize the envelope.
Even if the total energy deposited by the companion does not exceed the binding energy of the envelope, over a large volume the deposited energy can dominate. For example, the deposited energy can inflate the envelope. These are strongly interacting binary systems.

\emph{Angular momentum.} In the majority of binary systems, and in a large fraction of planetary systems, the orbital angular momentum is much larger than the spin angular momentum of each star. I consider a system to be strongly interacting by the angular momentum criterion if the amount of orbital angular momentum that is deposited to one (or two) of the stars is $J_{\rm dep} > \beta_j J_{\rm MS, max}$, where $J_{\rm MS, max}$ is the maximum angular momentum that the star can have on the main sequence (MS).
A star that along its entire evolution does not acquire angular momentum of that amount from a stellar or a substellar (brown dwarf or a planet) companion, is termed a Jsolated (for J-isolated) star (Sabach, E., \& Soker, N. 2017, arXiv:1704.05395)
\begin{equation}
J_{\rm dep} \le \beta_j J_{\rm MS, max} \qquad {\rm for~ a~ Jsolated~ star} .
\label{eq:jsoalted}
\end{equation}
At this point there is no accurate theory for the mass loss rate of red giant
stars, and we cannot determine the exact value of $\beta_j$. Efrat Sabach (talk in Nice) presented in her talk the class of Jsolated stars, and adopted values in the range $\beta_j \simeq 0.1-1$.

The assumption is that Jsolated stars have a mass loss rate much lower than the commonly used value. This is based on the claim that the empirical mass loss rate formulas are derived from observations of stellar populations that are contaminated by many binary systems and planetary systems. The companion, stellar or substellar, enhances the mass loss rate above the value that a Jsolated star would have.
In that respect we should note the comment made by
Nathan Smith (talk in ESO) that the derived mass loss rate from evolved massive stars was decreased by a factor of 3 due to clumping in the wind. Denise Keler (Nice) discussed the carbon AGB star CW Leo, and mentioned that its wind is clumpy.
Marie Van de Sande (Nice) discussed chemistry in clumpy AGB winds. The presence of clumps in AGB stars might imply that the derived mass loss rate was too high, and that the real mass loss rate of single AGB stars is lower than what is commonly assumed.
Carmen Sanchez Contreras (Nice) reported on higher mass loss rates than what regular evolutionary tracks give in some post-AGB stars. I attribute the higher mass loss rate to an interaction with a companion. Henri Boffin (Nice) mentioned that symbiotic primaries (the giant stars) have higher mass loss rates than regular giants. Interaction with the companion increases the mass loss rate.

\noindent \textbf{$\bigstar$ Summary of section 1.} There are several processes by which the secondary star can modify the evolution of the primary star. Some give rise to bright outbursts that are easily observed, other processes might be delicate. I argue that the deposition of angular momentum even by substellar objects can change the evolution of low mass RGB and AGB stars by increasing their mass loss rate. Stars isolated from acquiring angular momentum $\textbf{J}$, termed Jsolated stars, have much lower mass loss rates on their giant branches than usually assumed.

% ==========================================================
\section{FRACTION OF BINARY SYSTEMS}
\label{sec:fraction}
% ==========================================================

Max Moe (ESO) gave a very nice summary on the statistics of binary systems at formation. Missing was a discussion on the implications to late stellar evolutionary phases. For example, he concluded that most O stars suffer Roche lobe overflow (RLOF), which is a strong binary interaction. Hugues Sana (ESO) further raised the question on the binary distribution in stars of $8-15 M_\odot$ that are progenitors of most core collapse supernovae. My question following these talks was whether all core collapse supernovae suffered strong binary interaction? More extremely, is the phenomena of core collapse supernovae a binary one?
Nathan Smith (ESO) expressed his view that at least all stripped core collapse supernovae (those without hydrogen or without hydrogen and helium) are due to binary interaction. Schuyler Van Dyk (ESO), for example, presented a binary model for the Type IIb supernova 2016gkg.

Max Moe concluded also that only $ 15 \%$ of solar like stars experience RLOF. My input here was that many solar like stars can experience strong interaction with a substellar (a brown dwarf or a planet) companion. Ward Homan (Nice), for example, discussed the evolved giant L2 Puppis and concluded that low mass companions may play an important role in the morphological evolution of AGB stars. Eva Villaver (Nice) discussed the influence of planets on planetary nebulae (PNe).

Jen Winters (ESO) presented the finding according to which $26 \%$ of M stars are in binary systems, but did not study brown dwarf companions. Jay Fahiri then complained and asked ``who will do the physics of brown dwarf companions?''. My answer was that part of the physics was already done in a long study over the years (e.g., Soker, N. 1986, ApJ, 460, L53) of how brown dwarfs and planets influence the evolution of stars . Efrat Sabach in her poster (ESO) and talk (Nice) presented the class of Jasolated stars (see equation \ref{eq:jsoalted}), and how brown dwarfs and planets can turn a star to be non-Jsolated, hence shape the wind, leading to an elliptical PN. The chapter on the influence of substellar objects on stellar evolution has been already written for the `new textbook'.
%%% For example, In 1993 (Soker, N. 1993, "A definition of a binary system", in Luminous High-Latitude Stars, 45, 415): “I suggest that an appropriate definition of a binary system in the context of evolved stars is: “A star will be said to be in a binary system, if during some segment of its evolution, at least one of its properties is determined by an orbiting gravitationally bound object.” Note that this `object' can be . . . a brown dwarf or a planet.”
In any case, Pavel Kroupa (ESO) did present a study of brown dwarf companions, and pointed out that brown dwarfs have their own initial mass function.

Triple systems were also mentioned in the meetings. Silvia Toonen (ESO) discussed some systems that might be explained by triple-star evolution. This might be true for some of the systems she mentioned, but my view is that neither the outbursts of Eta Carinae nor the progenitor of SN Ia require any tertiary star in the system. Shlomi Hillel and Ron Schreier (ESO) presented a poster where they simulated the outcome of a merger of two low mass main sequence stars or brown dwarfs inside the envelope of an AGB star. This process, as well as other triple star interactions, can lead to the formation of `messy PNe'.

\noindent \textbf{$\bigstar$ Summary of section 2.} The new statistical studies that were presented suggest that massive stars (O and B stars on the main sequence) that do not interact with a binary companion end their life as one type or another of a `peculiar' astrophysical object. The fraction of low mass stars in interacting stellar binary systems is much lower, but many of them interact with substellar objects (brown dwarfs and planetary systems).

% ==========================================================
\section{MASS TRANSFER}
\label{sec:masstransfer}
% ==========================================================

Mass transfer is a major process in the evolution of binary systems, from birth, as reviewed by Robert Mathieu (ESO), and into their remnants, i.e., white dwarfs (WDs), neutron stars, or black holes. Young systems teach us that main sequence stars can accrete mass and launch jets.
I argue that main sequence stars can launch jets in the formation of barium (and similar) stars, when in a binary system with AGB and with post-AGB stars, some of which are progenitors of PNe, in some intermediate luminosity optical transients (ILOTS), and in many more types of binary systems.

Henri Boffin (ESO) gave an interesting and an informative opening talk on the types of binaries and on the types of their interactions (tidal, wind accretion, RLOF, common envelope). In one of his slides he stated that ``textbooks need most likely to be rewritten!'' However, in listing the different types of interactions he skipped the newly proposed grazing envelope evolution (GEE; presented by Sagiv Shiber in ESO and Nice). In the GEE the secondary star orbits at the outskirts of the envelope of the primary star and accretes mass through an accretion disk. The accretion disk launches jets that efficiently remove the envelope gas from the vicinity of the secondary star. The GEE can be viewed as a process where the companion continues to just enter a common envelope phase, but does not do it (at least for some time) because the jets remove the entire outer envelope. I will return to the grazing envelope evolution later on.

Another `new textbook chapter' that has already been written is on luminous blue variables (LBVs). Boffin stated on one slide ``. . .LBV . . we need to think more in terms of binary interaction". But as Amit Kashi (ESO) described in his talk, for several years he has been pointing out that all LBV major outbursts must be powered by binary interaction, where most of the energy comes from accretion of mass onto a companion.
Christophe Martayan (ESO) and Andrea Mehner (ESO) also raised the possibility that all LBVs are binaries.
Nathan Smith (Nice) talked about LBV and binarity, and argued that LBVs cannot be the progenitors of WR stars. He also claimed that recombination supplies the photons of the LBV outburst, as recombination energy escapes (see my comments on the common envelope evolution later on).
Nino Kochiashvili (ESO) studied the binarity of P Cygni (also Sopia Beradze; ESO), and found a number of periodicities in the light curve, which could be related to the presence of a binary companion. I note here that Amit Kashi in a paper from 2010 proposed that P Cygni is a binary system with a period of about 7 years. All these LBV binaries have high eccentricity, as IRC+10216 has (ESO poster by by Michael Bremer).

Amit Kashi (ESO and Nice) described how accretion onto the secondary star of Eta Carinae during periastron passages explains the behavior of Eta Carinae, and can account for its behavior during the two eruptions in the nineteenth century (there were other talks and posters on Eta Carinae, e.g., Joel Sanchez Bermudez, Gerd Weigelt; Noel Richardson).
Kashi also presented his claim for more massive stars in Eta Carinae, $170 M_\odot$ and $80 M_\odot$ for the LBV and companion, respectively, compared with the canonical masses of $120 M_\odot$ and $30 M_\odot$ .
His claim is based on dynamical fits of the orbit during the eruptions of the system, on behavior of He I lines in the last two spectroscopic events, and on recent detailed simulations that showed accretion close to periastron and its properties.

Tomasz Kaminski (ESO) presented an interesting review of some ILOTs that are mergers of two stars (are termed Red Novae, although they are not novae and are not always red). Amit Kashi and I expressed our view that these objects are related to many other types of binary systems, like Eta Carinae and other LBVs, some planetary nebulae, and more. Mass transfer occurs in all these outbursting stars that are termed ILOTs. A merger process is an extreme case of mass transfer, where one star transfers all (or most) of its mass in a short time.
See out ILOT page: https://phsites.technion.ac.il/soker/ilot-club/

There were many talks and posters on mass transfer and angular momentum transfer in massive binaries  beside LBVs.
Tomer Shenar (Nice) had an interesting talk where he argued that massive stars reach more or less the same point in the HR diagram where WR stars reside, whether they are single stars or in binary systems.
Avishai Gilkis (ESO and Nice) simulated the evolution of angular momentum in massive stars (ESO), and with a hydrodynamical code studied the collapse of a core of a massive star toward an explosion as a core collapse supernova. He finds that binary interactions are crucial for explosions of energetic supernovae and those with jets and magnetars, as they require fast rotation. Furthermore, due to non-spherical mass ejection during the binary interaction, likely a common envelope evolution phase, his results support   a strong connection between super-energetic supernovae (the jets) and their interaction with a bipolar massive circumstellar medium (formed during the binary interaction).
Florentin Millour (Nice) discussed Be stars, and Elizabeth Bartlett (Nice) discussed binary B[e] stars, some that  have polar outflows.

Hongwei Ge (ESO) presented an interesting result in a poster, according to which  the mass transfer from red giant branch (RGB) and AGB stars to a companion is more stable than what was thought before. I take this to imply a higher probability for launching jets, including for the GEE, as the companion can stay a longer time outside the primary envelope.

The so called process of wind-Roche lobe over flow was discussed in several talks and posters. In this process the mass losing star does not overflow its Roche lobe. However, the accretion rate is higher than that from a pure wind (for simulations of the wind-RLOF see, e.g., Mohamed, S., \& Podsiadlowski, P. 2007, Baltic Astronomy 16, 26 and 2012, Baltic Astronomy, 21, 88).
I see this process as accretion from an extended atmosphere as we proposed twenty years ago
(Harpaz, A., Rappaport, S., \& Soker, N. 1997, ApJ, 487, 809). In that paper we wrote: `` . . .if
the companion star is sufficiently close that the Roche lobe of the AGB star moves inside the extended atmosphere, then the slowly moving material will be forced to flow approximately along the critical potential surface (i.e., the Roche lobe) until it flows into the potential lobe of the companion star.''
In a later paper I termed this extended atmosphere envelope an effervescent zone (Soker, N. 2008, New Astronomy, 13, 491).
Gioia Rau (Nice), for example, made a dynamical model atmosphere for Mira stars with extended atmospheres.

Onno Pols (ESO) reviewed chemically polluted stars in binary systems (also Lewis Whitehouse and Ana Escorza). The polluted stars acquire their metals by accreting mass from their companion when it was on the AGB. Pols discussed the orbital separation of polluted stars, referring to the expectation of the traditional binary evolution calculation that during the mass transfer phase short binary systems will get shorter by the common envelope evolution, while wide binary systems will get wider by mass loss. Pols then mentioned that many chemically polluted  post-AGB binaries are right in the expected gap of the orbital separation.
This brings us to the next topic of the grazing envelope evolution (GEE).

\noindent \textbf{$\bigstar$ Summary of section 3.} Many types of erupting and exploding stars (LBVs and ILOTs in general; stripped supernovae; superluminous supernovae) require binary interaction, either for mass transfer or for angular momentum transfer. In many of these cases the accretion takes place through an accretion disk that launches jets. It seems that all major outbursts of LBVs are accompanied by jets, as well as the stripping of massive stars from their envelope.

% ==========================================================
\section{JETS IN POST-AGBIB STARS}
\label{sec:jets}
% ==========================================================

Post-AGBIBs, for post-AGB intermediate binaries, are binary systems composed of a post-AGB stars with a main sequence or a WD companion, at an orbital separation of $\approx 1 \AU$. Traditional evolutionary calculations predict either wider orbital separations because of mass loss, or closer post-AGB binaries because of a common envelope evolution. These systems received quite an attention in the two meetings.

Martha Irene Saladino (ESO and Nice) presented calculations of the evolution of the orbital separation in binaries with a primary AGB star. She included in her simulations mass accretion from the wind of the AGB star on to the companion. For slow winds, $V \la 15 \km \s^{-1}$, an accretion disk is formed around the companion. For slow winds (relative to orbital velocity) the wind carries a relatively large amount of angular momentum, and more systems go through a common envelope evolution. This causes the orbit to shrink. According to a comment by Hans Van Winckel the shrinkage of the orbit obtained is more than observed.

Hans Van Winckel (ESO) emphasized the presence of post-AGBIB in the `period gap' of traditional (old) binary models. The key point of his talk was that his group observes jets in many post-AGBIBs.
Moreover, the jets are launched by the companion to the post-AGB star and tend to have wide opening angles, as Dylan Bollen (Nice) discussed in his talk on the properties of jets in the post-AGBIB BD~$+46^\circ$442. I note here that Muhammad Akashi (Nice) presented new 3D hydrodynamical simulation of the shaping of PNe with wide jets, as he has been doing for several years now. The community should get used to the notion that in many cases jets are not well collimated, but rather they might be very wide, up to a half opening angle of about $70^\circ$.
Van Winckel wrote in his summary: ``Oribits are not explained (orbits, high eccentricities) (also poster by Joris Vos, Glenn-Michael Oomen; talk by Onno Pols, Ana Escorza, Devika Kamath).''
I think the GEE that I proposed in 2015, that is based on jets launched by the companion, can explain both the presence of post-AGBIBs in the `old gap' and the high eccentricities.

Before turning to the GEE, let me mention the important presentations of Devika Kamath (ESO and Nice). She summarized the group of post-RGBIBs (post-RGB intermediate binaries). They find them to be descendants of stars with zero age main sequence mass of $0.9-1.85 M_\odot$, that end with a mass of $0.25-0.48 M_\odot$. Some might experience mergers, some end as binary systems.
Although there is not yet proof of binarity in these systems, her talks were important in reminding us that many binary processes on the RGB are similar to those on the AGB. Todd Hillwig (Nice), for example,  mentioned that post-RGB stars can ionize ejected nebulae as well.

\begin{figure} [ht] %[h!]
\centering
\vskip -2.00 cm
\hskip -2.99 cm
 \includegraphics*[scale=0.85]{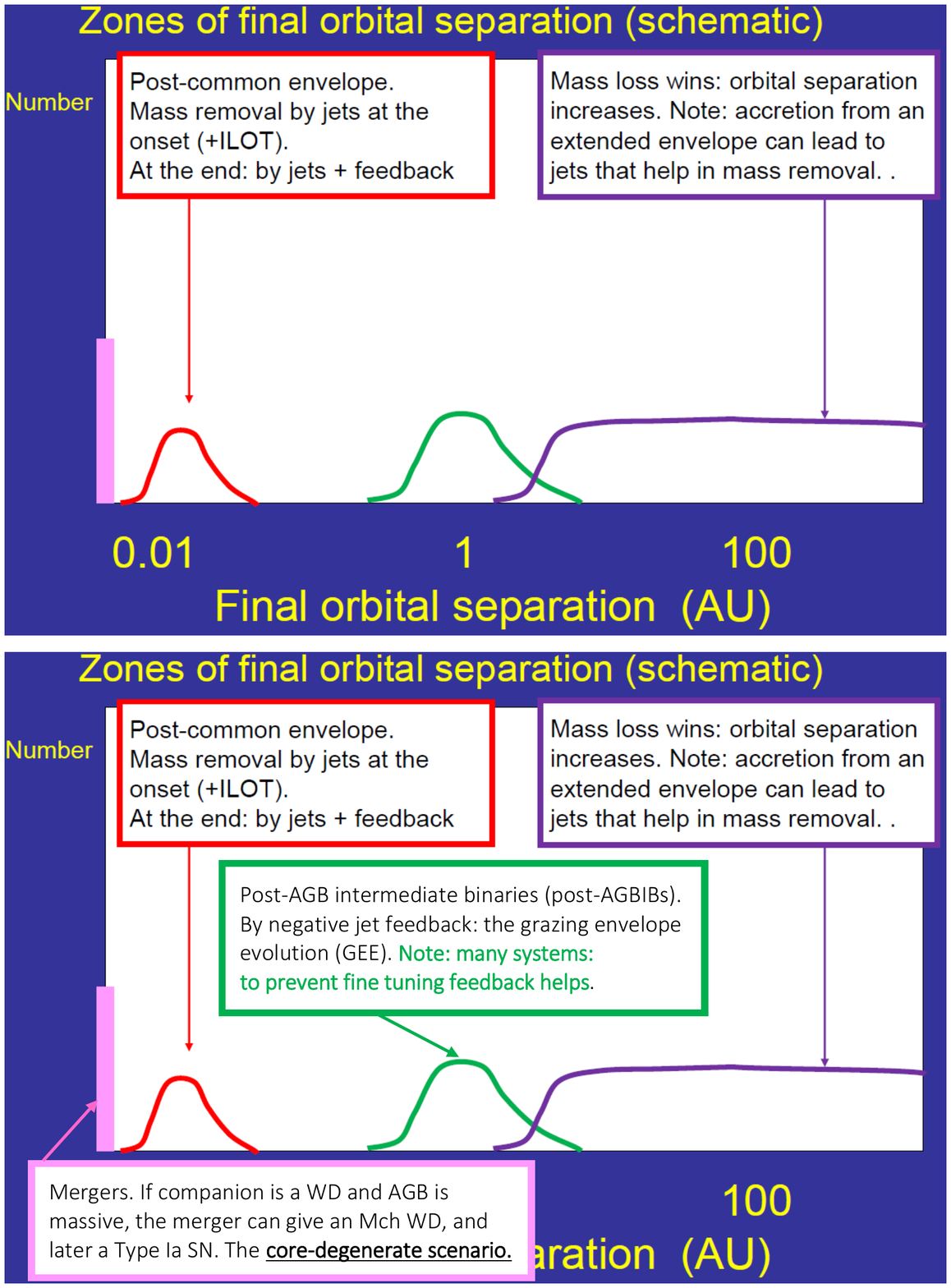} \\
\vskip -2.7 cm
\caption{A schematic presentation of the distribution of binary systems of post-RGB and post-AGB stars.
The upper panel schematically depicts the distribution of orbital separations that is expected in (old) traditional binary calculations. The green line depicts observed distribution that is unexplained by old binary calculations. In the lower panel the expectation of the GEE is drawn in green. In addition, in many cases the companion merges with the core, in particular for massive giants. One outcome might be a CO WD that is close to the Chandrasekhar mass, $M_{\rm Ch}$, and later explodes as SN Ia. This is the core degenerate scenario. Slides from my talk at Nice.  }
    \label{fig:schem}
\end{figure}

I attribute the intermediate orbital separation of post-AGBIBs, i.e., those systems that are in the `old gap' that traditional numerical simulations could not account for, to the GEE. This is presented schematically in figure \ref{fig:schem}. The jets that the companion launches efficiently remove the outer envelope of the giant star (an RGB or AGB star), to the degree that the system does not enter a common envelope, and hence the orbit does not shrink much (or not at all). Sagiv Shiber (a poster in ESO and a talk in Nice) presented 3D hydrodynamical simulations of the GEE that show the outer envelope removal.
In a recent paper I compared the formation of post-AGBIBs to the formation of the progenitors of Type IIb core-collapse supernovae, and presented my view that both types of systems experience the GEE (Soker, N. 2017, MNRAS, 470, L102).
% FFFFFFFFFFFFFFFFFFFFFFFFFFFFFFFFFFFFFFFFFFFFFFFF
% FFFFFFFFFFFFFFFFFFFFFFFFFFFFFFFFFFFFFFFFFFFFFFFF

Hans Van Winckel commented also on the high eccentricity of the post-AGBIBs. I pointed out that the binary system Eta Carinae has a very high eccentricity, and we know that it kept its period of about 5.5 years and that it launched jets in the Great Eruption.
Gloria Koenigsberger (ESO) mentioned in a poster a strong interaction near periastron. Such a strong interaction is required in the GEE explanation of the high eccentric orbits of post-AGBIBs.

Post-AGBIBs have circumbinary disks around them, e.g., as discussed by Valentin Bujarrabal (Nice). As mentioned in a poster by Lionel Siess (ESO), the circumbinary disc influences the evolution of the  orbital parameters. Of course, tidal forces and mass loss from the two stars also influence the orbital parameters. Rajeev Manick (Nice) presented two post-RGB binaries and four post-AGBIBs, all of which have a circumbinary disk and are eccentric.

There were more talks on jets.
Jeno Sokoloski (ESO) reviewed the class of symbiotic stars, and mentioned that many of them are observed to have jets; also Eric Lagadec (Nice) on precessing jets in the symbiotic star R~Aqr. From my perspective this is another example, in addition to post-AGBIB stars, that when it is possible, jets are observed. In many other cases where I argue we have jets, they are obscured: inside exploding stars, inside a common envelope, when there is lots of dust, e.g., as might be in the GEE.
She also mentioned that orbital periods of some symbiotic stars are shorter than expected by traditional binary calculations. Again, I attribute this to GEE.
Liz Bartlett (ESO) presented in a poster another example of polar outflow (jets) from interacting binaries.

\noindent \textbf{$\bigstar$ Summary of section 4.} Many systems present jets when the companion is outside and close to the surface of a giant star. Some of the jets are very wide. The community should get used to consider wide jets, with half opening angles of many tens of degrees.
The jets not only shape and accelerate the circumstellar gas, but they can also remove mass directly from the outskirts of the envelope. When they do it efficiently enough, the binary system enters the GEE. The GEE can explain many puzzles of traditional old calculations. One such puzzle involves the orbital separations of post-AGBIBs.

% ==========================================================
\section{COMMON ENVELOPE}
\label{sec:commonenvelope}
% ==========================================================

Orsola De Marco (Nice) showed new hydrodynamical simulations of the common envelope phase  that come closer to observations. However, the problems of envelope removal and the too large final orbital separation are not resolved yet even with these simulations.

David Jones (ESO) and Paulina Sowicka (ESO) discussed post-common envelope central binary systems in PNe. Jones focused on the common envelope evolution. He mentioned that when the companion is a main sequence star there are evidences that it is inflated in post-common envelope binaries, and this points that the companion accreted mass from the giant envelope. I argued that this goes along with the expectation when the companion launches jets, before, during, or after the common envelope phase.

Todd Hillwig (Nice) discussed the finding of a perfect alignment between binary axis and PN symmetry axis in some systems. He finds that about $12-15$ per cents of central binaries of PNe are WD-WD binaries. He argued for more observed CO-CO WDs than expected from numerical simulations; the simulations predict more He-CO WD binaries.

Veronika Schaffenroth (ESO) discussed the influence of sub-stellar objects on late stellar evolution, in particular via a common envelope evolution.
As was also pointed out by Efrat Sabach (ESO), the influence of brown dwarfs and planets on late stellar evolution must be taken into account.

Sebastian Ohlmann (ESO and Nice) presneted very nice 3D hydrodynamical simulations of the common envelope evolution (he is on the right track) and  a light curve based on the recombination energy. However, he used the same recombination energy to help unbind the envelope. At the meeting our group, Efrat Sabach, Shlomi Hillel, Ron Schreier, and I, expressed an agreement with Ohlmann that the recombination energy contributes significantly to the radiated energy, but we mentioned our result that the recombination energy has limited contribution to the ejection of the common envelope. We attribute a larger role to jets than to the recombination energy in removing the envelope (Sabach, E., Hillel, S., Schreier, R., Soker, N. 2017, arXiv:1706.05838).
As well, Norbert Langer (ESO) presented massive stars that evolved through the common envelope phase. To me this seems to be a problem to the recombination energy, as recombination energy is not large enough to eject envelopes of massive stars.

Robert Izzard (ESO and Nice) reviewed binary population synthesis, and emphasized that one of the most crucial uncertainties in performing binary population synthesis is the common envelope phase. Martyna Chruslinska (ESO) in her poster found common envelope evolution to be the major contribution to the uncertainty in the merger rates of two neutron stars.

\noindent \textbf{$\bigstar$ Summary of section 5.} The common envelope evolution is not solved yet, neither for stellar companions nor for planets that enter the envelope of a giant star. My view is that jets might play a crucial role. The companion might enter the envelope with already active jets. As I mentioned in previous sections, when the jets efficiently remove the outer envelope the companion will experience the GEE and will avoid the common envelope phase, at least at early stages. It is not clear yet whether the companion can launch jets during the common envelope phase. However, in a recent paper (Soker, N. 2017, arXiv:1706.03720) I argued that the companion can resume the jets activity when it exits the envelope from inside, near the core of the giant star. In most of these cases the jets operate in a negative feedback mechanism. Namely, too powerful jets clear gas from their surroundings and hence reduce the mass accretion rate. This in turn reduces the jet power. The negative feedback removes the need for a fine tuning.

% ==========================================================
\section{SHAPING THE OUTFLOW}
\label{sec:shaping}
% ==========================================================

Talks and posters on nebulae around evolved stars that were presented in the two meetings strengthened my (already strong) view that jets are behind the main shaping mechanism of these nebulae. I list several talks and posters below.

The high quality images that Claudia Agliozzo (ESO and Nice) presented of the LBVs HR~Carinae and RMC~127, and her analysis of the nebulae, clearly point to shaping by jets. She even commented on precessing jets, as expected in many cases of binary interaction.

Norbert Langer (ESO) Mentioned HD 148937, an O star with a bipolar nebula, and argued that this is a signature of a merger. I note that the progenitor of SN 1987A had a bipolar nebula, with the three pronounced rings, and it also went through a merger process.

Javier Alcolea (Nice) presented ALMA observations of OH231.8+4.2 (The Rotten Egg Nebula), and argued that the nebula was formed by three different mass loss events that have different centers. He implied that the  mass loss is done by an object that moves at a velocity of at least $6.5 \km \s^{-1}$, possibly an orbital motion. There was a discussion on why one lobe is longer than the other; this is indeed still a puzzle.

Lam Doan (Nice) presented the interacting winds shaping model of Frank \& Mellema (1994), and  a magnetic field shaping, and then a binary shaping. I note (as I did in several old papers) that these are not 3 different classes. For both interacting winds shaping and magnetic fields a binary interaction is necessary to supply the asymmetric mass loss and the angular momentum for the dynamo, respectively. The interacting winds model for shaping has several problems, e.g., it cannot explain precession seen in point-symmetric nebula, and I prefer shaping by jets.

Pierre Kervella (Nice) presented very impressive images and understandings of L2 Puppis. There is a disk with a radius of about $6 \AU$ seen in molecules. In the inner part the disk is Keplerian, while in its outer part it is sub-Keplerian. The central star, $M=0.653 M_\odot$, $R=123 R_\odot$, and $L=2000 L_\odot$, has a substellar companion of about $10-20$ Jupiter masses, i.e., the star interacts with its planetary system.

A hot topic in the two meeting was the formation of spiral patterns in the equatorial outflow from interacting binary systems. Eric Lagadec (Nice) presented spiral patterns around AGB stars. He mentioned that these patterns are clear indication for an orbital motion, and can result by modulation of the mass loss rate, winds collision, or gravitational influence by the companion. Elias Aydi (Nice) presented simulations that include pulsation, dust formation, and substellar objects. In these simulations when the orbital to pulsation period ratio is close to 2 spiral patterns appear.

Muhammad Akashi (Nice) presented new simulations of jets launched into a shell of an AGB wind. He obtained  H-shaped and barrel-shaped nebulae. The jets half-opening angle was $30^{\circ}$. His simulations over the years have shown the rich variety of morphologies that jets (from narrow to very wide jets) can form.

\noindent \textbf{$\bigstar$ Summary of section 6.} Jets are responsible for the shaping of the winds,  circumstellar matter, nebulae, and even explosions, in many stellar objects, in particular at their death. The source of the kinetic energy of the jets is accretion of mass onto the companion star, or onto the newly born neutron star in exploding massive stars. The jets energize and shape the outflow. Some morphological features are common to many types of nebulae shaped by jets: bubbles, ears, compressed equatorial gas, and many more.

% ==========================================================
\section{TYPE Ia SUPERNOVAE}
\label{sec:TypeIa}
% ==========================================================

Ferdinando Patat (ESO) reviewed the subject of SN Ia. I found his talk interesting in pointing the problems of the single degenerate and double degenerate scenarios, but I did not like his talk for not mentioning  the core-degenerate scenario. This omission of relevant scenarios from studies of SN Ia is an infectious disease that harms many papers, talks, and posters.
Carlo Abate (Nice) mentioned only the double degenerate and single degenerate scenarios for SN Ia. He argued that more mass accretion with wind-RLOF will increase the orbital separation from where these binary systems can form, hence bringing theory closer to observations.
However, as long as there is no consensus on the scenario of SN Ia, all scenarios should be mentioned. These include the core-degenerate scenario (which I think fits the best observations), the double degenerate scenario, the double detonation scenario, the single degenerate scenario, and the WD-WD collision scenario.

Chris Pritchet (ESO) mentioned 4 SN Ia scenarios (he omitted the WD-WD collision, that is the most problematic and cannot account for even $1 \%$  of all SN Ia). His aim was to distinguish between them by the delay time distribution (DTD) of SN Ia. The double degenerate scenario predicts $t^{-1}$ DTD, but this is below his observations. He used a new method that does not depend on star formation history. His conclusion was that the DTD goes more like $t^{-1.5}$ without cutoff. He excluded at 90\% the $t^{-1}$. This is somewhat a problem for the double-degenerate scenario.

Na'ama Hallakoun (ESO) described her search for double-WD population characteristics, and her finding that there are enough WD-WD systems to explain the SN Ia rate in the double degenerate scenario. I find that most of them are of too low mass to give the required mass to form manganese (Seitenzahl et al, 2013), or to give the  peak in masses of SN Ia at $1.4M_\odot$ (Scalzo, Ruiter \& Sim 2014).

There are also peculiar SN Ia. Ashley Ruiter (ESO) talked about electron-positron annihilation, and found that the merger of a He-WD and a CO-WD can explain SN~91bg-like events, that are less luminous than regular SN Ia.

\noindent \textbf{$\bigstar$ Summary of section 7.} When studying SN Ia progenitors we must distinguish between the most popular scenarios in the literature and the scenarios that might be the most popular among exploding WDs. Although the single degenerate and double degenerate are the most popular scenarios in our literature, I think that the core degenerate scenario is the most popular one among exploding WDs of regular SN Ia.

% ==========================================================
\section{SUMMARY: THE ALREADY WRITTEN CHAPTERS OF THE NEW TEXTBOOK}
\label{sec:summary}
% ==========================================================

As I already mentioned, the opening and closing talks of the ESO meeting referred to a `new textbook' that needs to be written. In the two meetings I argued that some chapters have already been written in recent years. I close my summary of the meetings by listing these `new chapters'.

\textbf{The chapter on intermediate luminosity optical transients (ILOTs).} In both meetings there was almost a consensus that LBVs are binary systems. Amit Kashi and I have been claiming that all major outbursts of LBVs are powered by mass transfer, i.e., gravitational energy released by mass transfer. An extreme event that is powered by mass transfer is the destruction of one star onto another, as is discussed for Red Novae (this really is a bad name). The connection of LBV major outbursts to merger events of main sequence (or slightly evolved) stars, has been written and teaches us a lot (see: https://phsites.technion.ac.il/soker/ilot-club/ ).

\textbf{The chapter on the grazing envelope evolution (GEE).} The main issue in the GEE is that jets remove the outer layer of the envelope in a negative feedback mechanism that might maintain a large orbital separation (see Fig. \ref{fig:schem}). In the GEE the jets might also remove mass from the extended envelope of the giant primary star, before the secondary star that launches the jets grazes the giant envelope. I think that the GEE solves some puzzles, such as the existence of post-AGBIBs and the very little hydrogen mass in Type IIb core collapse supernovae, and it is a very important new chapter in the `new textbook' (see: http://adsabs.harvard.edu/abs/2017MNRAS.470L.102S ).

\textbf{The chapter on jets in the common envelope evolution.} Jets can also help in removing the envelope during the common envelope phase. The operation of the jets via a negative feedback mechanism in the GEE and during common envelope evolution is related to other astrophysical objects, such as clusters of galaxies, where jets operate in a negative feedback mechanism (for a review see: http://adsabs.harvard.edu/abs/2016NewAR..75....1S )

\textbf{The chapter in the making on type Ia supernovae.} It is not clear yet what the progenitors of SN Ia are. As long as this puzzle is not solved, we should refer to all 5 binary scenarios. Although I prefer the core degenerate scenario, I admit that there are many open questions and problems with this scenario. The presence of ears in some type Ia supernova remnants suggests that jets play a role before the explosion, in shaping the circumstellar matter, or, less likely, during their explosion. (For more on ears and SN Ia scenarios: http://adsabs.harvard.edu/abs/2015MNRAS.447.2568T ).

\textbf{The chapter in the making on core collapse supernovae.} Most (or even all) progenitors of core collapse supernovae interact with a companion. Some progenitors require the companion star to strip their hydrogen/helium envelope and/or to spin-up their core. A rapidly rotating core is required for the formation of some superluminous supernovae, in particular those powered by a magnetar. The jets operate in a negative feedback mechanism that determines the outcome (see:
http://adsabs.harvard.edu/abs/2016ApJ...826..178G ).

\textbf{The chapter in the making on Jsolated stars.} The mass loss rate of truly single stars might be lower than what is commonly used. By truly single stars I refer to those that have no strong binary interaction, neither with stellar nor with substellar objects. Such stars are termed Jsolated stars, and are defined according to the angular momentum criterion in equation (\ref{eq:jsoalted}). Low mass Jsolated stars reach larger radii on the AGB, and by that increase their probability to interact with their planetary system (if they have one). This leads to elliptical PNe. Jsolated stars also reach higher luminosities during their post-AGB evolution. This might solve the puzzle of bright PNe in old stellar populations. (For more see: http://adsabs.harvard.edu/abs/2017arXiv170405395S )

I thank Henri Buffin for his every day help in putting together my \emph{live poster} during the ESO meeting. I thank my collaborators that have received with understanding my obsession with jets:  Muhammad Akashi, Efrat Sabach, Amit Kashi, Ron Schreier, Shlomi Hillel, Avishai Gilkis, Sagiv Shiber, Ealeal Bear, and Aldana Grichener.

% FFFFFFFFFFFFFFFFFFFFFFFFFFFFFFFFFFFFFFFFFFFFFFFF
%%% \begin{figure}[h!]
%%%   \centering
%%%    \includegraphics*[scale=0.55]{PapishGilkisSokerfigure1.eps} \\
%%%\caption{ A }
%%%      \label{fig:schem}
%%%\end{figure}
% FFFFFFFFFFFFFFFFFFFFFFFFFFFFFFFFFFFFFFFFFFFFFFFF

\label{lastpage}

\end{document}